\documentclass{ws-p8-50x6-00}

\begin{document}

\title{DIRAC Experiment and test of Low-Energy QCD}

\author{M. Pentia} 

\address{NIPNE-HH, P.O.Box MG-6, 76900, Bucharest-Magurele,
ROMANIA\\E-mail: pentia@ifin.nipne.ro \\
\vspace*{0.3cm}
on behalf of \\
"The DIRAC Collaboration"\\
\vspace*{0.3cm}
B.~Adeva$^o$, L.~Afanasev$^l$, M.~Benayoun$^d$, 
V.~Brekhovskikh$^n$,
G.~Caragheorgheopol$^m$,T.~Cechak$^b$, M.~Chiba$^j$, S.~Constantinescu$^m$,
A.~Doudarev$^l$, D.~Dreossi$^f$, D.~Drijard$^a$, 
M.~Ferro-Luzzi$^a$, T.~Gallas Torreira$^{a,o}$, J.~Gerndt$^b$, R.~Giacomich$^f$,
P.~Gianotti$^e$, F.~Gomez$^o$, A.~Gorin$^n$, O.~Gortchakov$^l$, 
C.~Guaraldo$^e$, M.~Hansroul$^a$, R.~Hosek$^b$,
M.~Iliescu$^{e,m}$, N.~Kalinina$^l$, V.~Karpoukhine$^l$, 
J.~Kluson$^b$, M.~Kobayashi$^g$, P.~Kokkas$^p$, 
V.~Komarov$^l$, A.~Koulikov$^l$, 
A.~Kouptsov$^l$, V.~Krouglov$^l$, L.~Krouglova$^l$, K.-I.~Kuroda$^k$,
A.~Lanaro$^{a,e}$, V.~Lapshine$^n$, R.~Lednicky$^c$, P.~Leruste$^d$, 
P.~Levisandri$^e$, A.~Lopez Aguera$^o$, V.~Lucherini$^e$, 
T.~Maki$^i$, I.~Manuilov$^n$, L.~Montanet$^a$, 
J.-L.~Narjoux$^d$, L.~Nemenov$^{a,l}$, M.~Nikitin$^l$, 
T.~Nunez Pardo$^o$, K.~Okada$^h$, V.~Olchevskii$^l$, 
A.~Pazos$^o$, M.~Pentia$^m$, A.~Penzo$^f$, J.-M.~Perreau$^a$, 
C.~Petrascu$^{e,m}$, M.~Plo$^o$, T.~Ponta$^m$, D.~Pop$^m$, 
A.~Riazantsev$^n$, J.M.~Rodriguez$^o$,
A.~Rodriguez Fernandez$^o$, V.~Rykaline$^n$,  
C.~Santamarina$^o$, J.~Schacher$^q$, A.~Sidorov$^n$, J.~Smolik$^c$,
F.~Takeutchi$^h$, A.~Tarasov$^l$, L.~Tauscher$^p$, S.~Trousov$^l$, 
P.~Vazquez$^o$, S.~Vlachos$^p$, V.~Yazkov$^l$, Y.~Yoshimura$^g$, P.~Zrelov$^l$ \\
\vspace*{0.3cm}
$^a$ CERN, Geneva, Switzerland ;  
$^b$ Czech Technical University, Prague, Czech Republic ; 
$^c$ Prague University, Czech Republic ;
$^d$ LPNHE des Universites Paris VI/VII, IN2P3-CNRS, France ;
$^e$ INFN - Laboratori Nazionali di Frascati, Frascati, Italy ;
$^f$ Trieste University and  INFN-Trieste, Italy ;
$^g$ KEK, Tsukuba, Japan ; 
$^h$ Kyoto Sangyou University, Japan ;
$^i$ UOEH-Kyushu, Japan ;
$^j$  Tokyo Metropolitan University, Japan ;
$^k$ Waseda University, Japan ;
$^l$ JINR Dubna, Russia ;
$^m$ National Institute for 
Physics and Nuclear Engineering NIPNE-HH, Bucharest, Romania ; 
$^n$ IHEP Protvino, Russia ;
$^o$ Santiago de Compostela University, Spain ; 
$^p$ Basel University, Switzerland  ;
$^q$ Bern University, Switzerland 
}

\maketitle

\abstracts{
The low-energy QCD predictions to be tested by the DIRAC experiment are revised.
The experimental method, the setup characteristics and capabilities, along with
first experimental results are reported. Preliminary analysis shows good 
detector performance: alignment error via $\Lambda$ mass measurement
$m_\Lambda = 1115.6~MeV/c^2$ with $\sigma = 0.92~MeV/c^2$, $p \pi^-$ relative
momentum resolution $\sigma_Q \approx 2.7~MeV/c$, and evidence for 
$\pi^+ \pi^-$ low momentum Coulomb correlation.
}

\vspace{-10mm}
\section{Introduction}
Quantum Chromodynamics (QCD), responsible for the strong interaction sector of 
the Standard Model (SM) has successfully been tested only in the perturbative 
region of high momentum transfer ($Q>1~GeV$) or at short relative distance 
$\Delta r \sim \hbar / Q$ ($\Delta r < 0.2 fm)$. Here the constituent quarks 
behave as weakly interacting, nearly massless particles. The QCD in the 
perturbative region, as any gauge theory with massless fermions, presents 
chiral symmetry.

In the nonperturbative region of low momentum transfer (low-energy), say 
$Q < 100~MeV$, or equivalently at large distance ($\Delta r > 2 fm$), 
asymptotic freedom is absent, and quark confinement takes place. In the low 
energy region the chiral symmetry of QCD must be spontaneously broken. 

The {\sl Chiral Perturbation Theory} (ChPT)\cite{Gasser84,Gasser85} 
seems to be the candidate theory for low energy processes. It exploits the 
mechanism of spontaneous breakdown of chiral symmetry (SBChS), or, in other
words, the existence of a quark condensate.
In order to test the existence of the quark condensate, the particularly 
signi\-ficant symmetry breaking effect refers to the $S$-{\sl wave $\pi \pi$ 
scattering lengths}. From the theoretical point of view, $\pi \pi$ scattering 
is a deeply studied problem. Within standard ChPT, Gasser and 
Leutwyler\cite{Gasser84,Gasser85,Gasser83} and also Bijnens and
collaborators\cite{Bijnens97} as well as within the {\sl Generalized Chiral 
Perturbation Theory} (GChPT), Stern and collaborators\cite{Stern93,Knecht95a},
have obtained expressions for the $\pi \pi$ scattering amplitude in the chiral 
expansion. 

The leading order expansion of the scattering amplitude is\cite{Knecht95a}
\begin{equation}\label{amplitude1}
A(s;t,u) = \alpha \frac{M_{\pi}^2}{3 F_\pi ^2} + \frac{\beta}{F_\pi ^2}
\left( s - \frac{4}{3} M_{\pi}^2 \right) + {\cal O}(p^4)
\end{equation}
where $\alpha$ and $\beta$ encode the information on the strength of the quark 
condensate. In the limit of a strong quark condensate, one has 
$\alpha \approx 1$, $\beta \approx 1$. A substantial departure of $\alpha$ from
unity signals a much smaller value of the condensate.

The values predicted for the isospin $I=0$ and $I=2$ $S$-wave scattering 
lengths $a_0^0$ and $a_0^2$ can be confronted with the future experimental 
values of the DIRAC experiment.
The available experimental data\cite{Rosselet77} for the scattering length 
is $a_0 ^0 = 0.26 \pm 0.05$. Based on these data there is no possibility to 
estimate the strength of the quark condensate and so to measure the extent of 
chiral symmetry breaking.

The DIRAC experiment\cite{Adeva94} aims to determine the difference of the 
scattering lengths $\Delta = |a_0 ^0 - a_0 ^2 |$ with 5\% accuracy, by 
measuring the lifetime of pionium ($\pi^+ \pi^-$ bound state). For the first 
time experimental evidence in favour of or against the existence of a strong 
quark condensate in the QCD vacuum could be within reach.

\vspace{-2mm}
\section{Pionium lifetime}
Pionium is a metastable bound state, produced by $\pi^+$ and $\pi^-$
electromagnetic interaction and decaying into $\pi^0 \pi^0$ due to strong
interaction. To obtain pion scattering lengths in a model independent way, a 
measurement of the lifetime of pionium has been proposed many years ago by
Nemenov\cite{Nemenov85}. The measurement of the pionium lifetime ($\tau$) will
allow to determine the difference $|a_0^0 - a_0^2|$ of the strong $S$-wave
$\pi \pi$-scattering lengths for isospin $I=0$ and $I=2$.

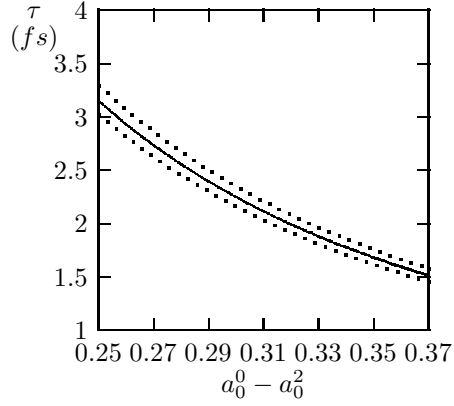
\begin{figure}[h]
\begin{center}
\setlength{\unitlength}{0.2pt}
\ifx\plotpoint\undefined\newsavebox{\plotpoint}\fi
\sbox{\plotpoint}{\rule[-0.200pt]{0.400pt}{0.400pt}}%
\begin{picture}(825,810)(0,0)
\font\gnuplot=cmr10 at 10pt
\gnuplot
\sbox{\plotpoint}{\rule[-0.200pt]{0.400pt}{0.400pt}}%
\put(181.0,163.0){\rule[-0.200pt]{4.818pt}{0.400pt}}
\put(161,163){\makebox(0,0)[r]{1}}
\put(785.0,163.0){\rule[-0.200pt]{4.818pt}{0.400pt}}
\put(181.0,264.0){\rule[-0.200pt]{4.818pt}{0.400pt}}
\put(161,264){\makebox(0,0)[r]{1.5}}
\put(785.0,264.0){\rule[-0.200pt]{4.818pt}{0.400pt}}
\put(181.0,365.0){\rule[-0.200pt]{4.818pt}{0.400pt}}
\put(161,365){\makebox(0,0)[r]{2}}
\put(785.0,365.0){\rule[-0.200pt]{4.818pt}{0.400pt}}
\put(181.0,466.0){\rule[-0.200pt]{4.818pt}{0.400pt}}
\put(161,466){\makebox(0,0)[r]{2.5}}
\put(785.0,466.0){\rule[-0.200pt]{4.818pt}{0.400pt}}
\put(181.0,567.0){\rule[-0.200pt]{4.818pt}{0.400pt}}
\put(161,567){\makebox(0,0)[r]{3}}
\put(785.0,567.0){\rule[-0.200pt]{4.818pt}{0.400pt}}
\put(181.0,668.0){\rule[-0.200pt]{4.818pt}{0.400pt}}
\put(161,668){\makebox(0,0)[r]{3.5}}
\put(785.0,668.0){\rule[-0.200pt]{4.818pt}{0.400pt}}
\put(181.0,769.0){\rule[-0.200pt]{4.818pt}{0.400pt}}
\put(161,769){\makebox(0,0)[r]{4}}
\put(785.0,769.0){\rule[-0.200pt]{4.818pt}{0.400pt}}
\put(181.0,163.0){\rule[-0.200pt]{0.400pt}{4.818pt}}
\put(181,122){\makebox(0,0){0.25}}
\put(181.0,749.0){\rule[-0.200pt]{0.400pt}{4.818pt}}
\put(285.0,163.0){\rule[-0.200pt]{0.400pt}{4.818pt}}
\put(285,122){\makebox(0,0){0.27}}
\put(285.0,749.0){\rule[-0.200pt]{0.400pt}{4.818pt}}
\put(389.0,163.0){\rule[-0.200pt]{0.400pt}{4.818pt}}
\put(389,122){\makebox(0,0){0.29}}
\put(389.0,749.0){\rule[-0.200pt]{0.400pt}{4.818pt}}
\put(493.0,163.0){\rule[-0.200pt]{0.400pt}{4.818pt}}
\put(493,122){\makebox(0,0){0.31}}
\put(493.0,749.0){\rule[-0.200pt]{0.400pt}{4.818pt}}
\put(597.0,163.0){\rule[-0.200pt]{0.400pt}{4.818pt}}
\put(597,122){\makebox(0,0){0.33}}
\put(597.0,749.0){\rule[-0.200pt]{0.400pt}{4.818pt}}
\put(701.0,163.0){\rule[-0.200pt]{0.400pt}{4.818pt}}
\put(701,122){\makebox(0,0){0.35}}
\put(701.0,749.0){\rule[-0.200pt]{0.400pt}{4.818pt}}
\put(805.0,163.0){\rule[-0.200pt]{0.400pt}{4.818pt}}
\put(805,122){\makebox(0,0){0.37}}
\put(805.0,749.0){\rule[-0.200pt]{0.400pt}{4.818pt}}

\put(181.0,163.0){\rule[-0.200pt]{120.322pt}{0.400pt}}

\put(805.0,163.0){\rule[-0.200pt]{0.400pt}{120.985pt}}

\put(181.0,769.0){\rule[-0.200pt]{120.322pt}{0.400pt}}

\put(60,766){\makebox(0,0){$\tau$}}
\put(60,716){\makebox(0,0){$(fs)$}}
\put(493,61){\makebox(0,0){$a_0^0-a_0^2$}}

\put(181.0,163.0){\rule[-0.200pt]{0.400pt}{120.985pt}}

\multiput(181.00,596.93)(0.599,-0.477){7}{\rule{0.580pt}{0.115pt}}
\multiput(181.00,597.17)(4.796,-5.000){2}{\rule{0.290pt}{0.400pt}}
\multiput(187.00,591.93)(0.581,-0.482){9}{\rule{0.567pt}{0.116pt}}
\multiput(187.00,592.17)(5.824,-6.000){2}{\rule{0.283pt}{0.400pt}}
\multiput(194.00,585.93)(0.599,-0.477){7}{\rule{0.580pt}{0.115pt}}
\multiput(194.00,586.17)(4.796,-5.000){2}{\rule{0.290pt}{0.400pt}}
\multiput(200.00,580.93)(0.491,-0.482){9}{\rule{0.500pt}{0.116pt}}
\multiput(200.00,581.17)(4.962,-6.000){2}{\rule{0.250pt}{0.400pt}}
\multiput(206.00,574.93)(0.710,-0.477){7}{\rule{0.660pt}{0.115pt}}
\multiput(206.00,575.17)(5.630,-5.000){2}{\rule{0.330pt}{0.400pt}}
\multiput(213.00,569.93)(0.599,-0.477){7}{\rule{0.580pt}{0.115pt}}
\multiput(213.00,570.17)(4.796,-5.000){2}{\rule{0.290pt}{0.400pt}}
\multiput(219.00,564.93)(0.491,-0.482){9}{\rule{0.500pt}{0.116pt}}
\multiput(219.00,565.17)(4.962,-6.000){2}{\rule{0.250pt}{0.400pt}}
\multiput(225.00,558.93)(0.599,-0.477){7}{\rule{0.580pt}{0.115pt}}
\multiput(225.00,559.17)(4.796,-5.000){2}{\rule{0.290pt}{0.400pt}}
\multiput(231.00,553.93)(0.710,-0.477){7}{\rule{0.660pt}{0.115pt}}
\multiput(231.00,554.17)(5.630,-5.000){2}{\rule{0.330pt}{0.400pt}}
\multiput(238.00,548.93)(0.599,-0.477){7}{\rule{0.580pt}{0.115pt}}
\multiput(238.00,549.17)(4.796,-5.000){2}{\rule{0.290pt}{0.400pt}}
\multiput(244.00,543.93)(0.599,-0.477){7}{\rule{0.580pt}{0.115pt}}
\multiput(244.00,544.17)(4.796,-5.000){2}{\rule{0.290pt}{0.400pt}}
\multiput(250.00,538.93)(0.710,-0.477){7}{\rule{0.660pt}{0.115pt}}
\multiput(250.00,539.17)(5.630,-5.000){2}{\rule{0.330pt}{0.400pt}}
\multiput(257.00,533.93)(0.599,-0.477){7}{\rule{0.580pt}{0.115pt}}
\multiput(257.00,534.17)(4.796,-5.000){2}{\rule{0.290pt}{0.400pt}}
\multiput(263.00,528.94)(0.774,-0.468){5}{\rule{0.700pt}{0.113pt}}
\multiput(263.00,529.17)(4.547,-4.000){2}{\rule{0.350pt}{0.400pt}}
\multiput(269.00,524.93)(0.710,-0.477){7}{\rule{0.660pt}{0.115pt}}
\multiput(269.00,525.17)(5.630,-5.000){2}{\rule{0.330pt}{0.400pt}}
\multiput(276.00,519.93)(0.599,-0.477){7}{\rule{0.580pt}{0.115pt}}
\multiput(276.00,520.17)(4.796,-5.000){2}{\rule{0.290pt}{0.400pt}}
\multiput(282.00,514.94)(0.774,-0.468){5}{\rule{0.700pt}{0.113pt}}
\multiput(282.00,515.17)(4.547,-4.000){2}{\rule{0.350pt}{0.400pt}}
\multiput(288.00,510.93)(0.599,-0.477){7}{\rule{0.580pt}{0.115pt}}
\multiput(288.00,511.17)(4.796,-5.000){2}{\rule{0.290pt}{0.400pt}}
\multiput(294.00,505.94)(0.920,-0.468){5}{\rule{0.800pt}{0.113pt}}
\multiput(294.00,506.17)(5.340,-4.000){2}{\rule{0.400pt}{0.400pt}}
\multiput(301.00,501.93)(0.599,-0.477){7}{\rule{0.580pt}{0.115pt}}
\multiput(301.00,502.17)(4.796,-5.000){2}{\rule{0.290pt}{0.400pt}}
\multiput(307.00,496.94)(0.774,-0.468){5}{\rule{0.700pt}{0.113pt}}
\multiput(307.00,497.17)(4.547,-4.000){2}{\rule{0.350pt}{0.400pt}}
\multiput(313.00,492.94)(0.920,-0.468){5}{\rule{0.800pt}{0.113pt}}
\multiput(313.00,493.17)(5.340,-4.000){2}{\rule{0.400pt}{0.400pt}}
\multiput(320.00,488.93)(0.599,-0.477){7}{\rule{0.580pt}{0.115pt}}
\multiput(320.00,489.17)(4.796,-5.000){2}{\rule{0.290pt}{0.400pt}}
\multiput(326.00,483.94)(0.774,-0.468){5}{\rule{0.700pt}{0.113pt}}
\multiput(326.00,484.17)(4.547,-4.000){2}{\rule{0.350pt}{0.400pt}}
\multiput(332.00,479.94)(0.920,-0.468){5}{\rule{0.800pt}{0.113pt}}
\multiput(332.00,480.17)(5.340,-4.000){2}{\rule{0.400pt}{0.400pt}}
\multiput(339.00,475.94)(0.774,-0.468){5}{\rule{0.700pt}{0.113pt}}
\multiput(339.00,476.17)(4.547,-4.000){2}{\rule{0.350pt}{0.400pt}}
\multiput(345.00,471.94)(0.774,-0.468){5}{\rule{0.700pt}{0.113pt}}
\multiput(345.00,472.17)(4.547,-4.000){2}{\rule{0.350pt}{0.400pt}}
\multiput(351.00,467.94)(0.920,-0.468){5}{\rule{0.800pt}{0.113pt}}
\multiput(351.00,468.17)(5.340,-4.000){2}{\rule{0.400pt}{0.400pt}}
\multiput(358.00,463.94)(0.774,-0.468){5}{\rule{0.700pt}{0.113pt}}
\multiput(358.00,464.17)(4.547,-4.000){2}{\rule{0.350pt}{0.400pt}}
\multiput(364.00,459.94)(0.774,-0.468){5}{\rule{0.700pt}{0.113pt}}
\multiput(364.00,460.17)(4.547,-4.000){2}{\rule{0.350pt}{0.400pt}}
\multiput(370.00,455.94)(0.774,-0.468){5}{\rule{0.700pt}{0.113pt}}
\multiput(370.00,456.17)(4.547,-4.000){2}{\rule{0.350pt}{0.400pt}}
\multiput(376.00,451.94)(0.920,-0.468){5}{\rule{0.800pt}{0.113pt}}
\multiput(376.00,452.17)(5.340,-4.000){2}{\rule{0.400pt}{0.400pt}}
\multiput(383.00,447.94)(0.774,-0.468){5}{\rule{0.700pt}{0.113pt}}
\multiput(383.00,448.17)(4.547,-4.000){2}{\rule{0.350pt}{0.400pt}}
\multiput(389.00,443.95)(1.132,-0.447){3}{\rule{0.900pt}{0.108pt}}
\multiput(389.00,444.17)(4.132,-3.000){2}{\rule{0.450pt}{0.400pt}}
\multiput(395.00,440.94)(0.920,-0.468){5}{\rule{0.800pt}{0.113pt}}
\multiput(395.00,441.17)(5.340,-4.000){2}{\rule{0.400pt}{0.400pt}}
\multiput(402.00,436.94)(0.774,-0.468){5}{\rule{0.700pt}{0.113pt}}
\multiput(402.00,437.17)(4.547,-4.000){2}{\rule{0.350pt}{0.400pt}}
\multiput(408.00,432.95)(1.132,-0.447){3}{\rule{0.900pt}{0.108pt}}
\multiput(408.00,433.17)(4.132,-3.000){2}{\rule{0.450pt}{0.400pt}}
\multiput(414.00,429.94)(0.920,-0.468){5}{\rule{0.800pt}{0.113pt}}
\multiput(414.00,430.17)(5.340,-4.000){2}{\rule{0.400pt}{0.400pt}}
\multiput(421.00,425.95)(1.132,-0.447){3}{\rule{0.900pt}{0.108pt}}
\multiput(421.00,426.17)(4.132,-3.000){2}{\rule{0.450pt}{0.400pt}}
\multiput(427.00,422.94)(0.774,-0.468){5}{\rule{0.700pt}{0.113pt}}
\multiput(427.00,423.17)(4.547,-4.000){2}{\rule{0.350pt}{0.400pt}}
\multiput(433.00,418.95)(1.132,-0.447){3}{\rule{0.900pt}{0.108pt}}
\multiput(433.00,419.17)(4.132,-3.000){2}{\rule{0.450pt}{0.400pt}}
\multiput(439.00,415.94)(0.920,-0.468){5}{\rule{0.800pt}{0.113pt}}
\multiput(439.00,416.17)(5.340,-4.000){2}{\rule{0.400pt}{0.400pt}}
\multiput(446.00,411.95)(1.132,-0.447){3}{\rule{0.900pt}{0.108pt}}
\multiput(446.00,412.17)(4.132,-3.000){2}{\rule{0.450pt}{0.400pt}}
\multiput(452.00,408.94)(0.774,-0.468){5}{\rule{0.700pt}{0.113pt}}
\multiput(452.00,409.17)(4.547,-4.000){2}{\rule{0.350pt}{0.400pt}}
\multiput(458.00,404.95)(1.355,-0.447){3}{\rule{1.033pt}{0.108pt}}
\multiput(458.00,405.17)(4.855,-3.000){2}{\rule{0.517pt}{0.400pt}}
\multiput(465.00,401.95)(1.132,-0.447){3}{\rule{0.900pt}{0.108pt}}
\multiput(465.00,402.17)(4.132,-3.000){2}{\rule{0.450pt}{0.400pt}}
\multiput(471.00,398.95)(1.132,-0.447){3}{\rule{0.900pt}{0.108pt}}
\multiput(471.00,399.17)(4.132,-3.000){2}{\rule{0.450pt}{0.400pt}}
\multiput(477.00,395.94)(0.920,-0.468){5}{\rule{0.800pt}{0.113pt}}
\multiput(477.00,396.17)(5.340,-4.000){2}{\rule{0.400pt}{0.400pt}}
\multiput(484.00,391.95)(1.132,-0.447){3}{\rule{0.900pt}{0.108pt}}
\multiput(484.00,392.17)(4.132,-3.000){2}{\rule{0.450pt}{0.400pt}}
\multiput(490.00,388.95)(1.132,-0.447){3}{\rule{0.900pt}{0.108pt}}
\multiput(490.00,389.17)(4.132,-3.000){2}{\rule{0.450pt}{0.400pt}}
\multiput(496.00,385.95)(1.355,-0.447){3}{\rule{1.033pt}{0.108pt}}
\multiput(496.00,386.17)(4.855,-3.000){2}{\rule{0.517pt}{0.400pt}}
\multiput(503.00,382.95)(1.132,-0.447){3}{\rule{0.900pt}{0.108pt}}
\multiput(503.00,383.17)(4.132,-3.000){2}{\rule{0.450pt}{0.400pt}}
\multiput(509.00,379.95)(1.132,-0.447){3}{\rule{0.900pt}{0.108pt}}
\multiput(509.00,380.17)(4.132,-3.000){2}{\rule{0.450pt}{0.400pt}}
\multiput(515.00,376.95)(1.132,-0.447){3}{\rule{0.900pt}{0.108pt}}
\multiput(515.00,377.17)(4.132,-3.000){2}{\rule{0.450pt}{0.400pt}}
\multiput(521.00,373.95)(1.355,-0.447){3}{\rule{1.033pt}{0.108pt}}
\multiput(521.00,374.17)(4.855,-3.000){2}{\rule{0.517pt}{0.400pt}}
\multiput(528.00,370.95)(1.132,-0.447){3}{\rule{0.900pt}{0.108pt}}
\multiput(528.00,371.17)(4.132,-3.000){2}{\rule{0.450pt}{0.400pt}}
\multiput(534.00,367.95)(1.132,-0.447){3}{\rule{0.900pt}{0.108pt}}
\multiput(534.00,368.17)(4.132,-3.000){2}{\rule{0.450pt}{0.400pt}}
\multiput(540.00,364.95)(1.355,-0.447){3}{\rule{1.033pt}{0.108pt}}
\multiput(540.00,365.17)(4.855,-3.000){2}{\rule{0.517pt}{0.400pt}}
\multiput(547.00,361.95)(1.132,-0.447){3}{\rule{0.900pt}{0.108pt}}
\multiput(547.00,362.17)(4.132,-3.000){2}{\rule{0.450pt}{0.400pt}}
\put(553,358.17){\rule{1.300pt}{0.400pt}}
\multiput(553.00,359.17)(3.302,-2.000){2}{\rule{0.650pt}{0.400pt}}
\multiput(559.00,356.95)(1.355,-0.447){3}{\rule{1.033pt}{0.108pt}}
\multiput(559.00,357.17)(4.855,-3.000){2}{\rule{0.517pt}{0.400pt}}
\multiput(566.00,353.95)(1.132,-0.447){3}{\rule{0.900pt}{0.108pt}}
\multiput(566.00,354.17)(4.132,-3.000){2}{\rule{0.450pt}{0.400pt}}
\multiput(572.00,350.95)(1.132,-0.447){3}{\rule{0.900pt}{0.108pt}}
\multiput(572.00,351.17)(4.132,-3.000){2}{\rule{0.450pt}{0.400pt}}
\put(578,347.17){\rule{1.500pt}{0.400pt}}
\multiput(578.00,348.17)(3.887,-2.000){2}{\rule{0.750pt}{0.400pt}}
\multiput(585.00,345.95)(1.132,-0.447){3}{\rule{0.900pt}{0.108pt}}
\multiput(585.00,346.17)(4.132,-3.000){2}{\rule{0.450pt}{0.400pt}}
\multiput(591.00,342.95)(1.132,-0.447){3}{\rule{0.900pt}{0.108pt}}
\multiput(591.00,343.17)(4.132,-3.000){2}{\rule{0.450pt}{0.400pt}}
\put(597,339.17){\rule{1.300pt}{0.400pt}}
\multiput(597.00,340.17)(3.302,-2.000){2}{\rule{0.650pt}{0.400pt}}
\multiput(603.00,337.95)(1.355,-0.447){3}{\rule{1.033pt}{0.108pt}}
\multiput(603.00,338.17)(4.855,-3.000){2}{\rule{0.517pt}{0.400pt}}
\multiput(610.00,334.95)(1.132,-0.447){3}{\rule{0.900pt}{0.108pt}}
\multiput(610.00,335.17)(4.132,-3.000){2}{\rule{0.450pt}{0.400pt}}
\put(616,331.17){\rule{1.300pt}{0.400pt}}
\multiput(616.00,332.17)(3.302,-2.000){2}{\rule{0.650pt}{0.400pt}}
\multiput(622.00,329.95)(1.355,-0.447){3}{\rule{1.033pt}{0.108pt}}
\multiput(622.00,330.17)(4.855,-3.000){2}{\rule{0.517pt}{0.400pt}}
\put(629,326.17){\rule{1.300pt}{0.400pt}}
\multiput(629.00,327.17)(3.302,-2.000){2}{\rule{0.650pt}{0.400pt}}
\multiput(635.00,324.95)(1.132,-0.447){3}{\rule{0.900pt}{0.108pt}}
\multiput(635.00,325.17)(4.132,-3.000){2}{\rule{0.450pt}{0.400pt}}
\put(641,321.17){\rule{1.500pt}{0.400pt}}
\multiput(641.00,322.17)(3.887,-2.000){2}{\rule{0.750pt}{0.400pt}}
\put(648,319.17){\rule{1.300pt}{0.400pt}}
\multiput(648.00,320.17)(3.302,-2.000){2}{\rule{0.650pt}{0.400pt}}
\multiput(654.00,317.95)(1.132,-0.447){3}{\rule{0.900pt}{0.108pt}}
\multiput(654.00,318.17)(4.132,-3.000){2}{\rule{0.450pt}{0.400pt}}
\put(660,314.17){\rule{1.300pt}{0.400pt}}
\multiput(660.00,315.17)(3.302,-2.000){2}{\rule{0.650pt}{0.400pt}}
\put(666,312.17){\rule{1.500pt}{0.400pt}}
\multiput(666.00,313.17)(3.887,-2.000){2}{\rule{0.750pt}{0.400pt}}
\multiput(673.00,310.95)(1.132,-0.447){3}{\rule{0.900pt}{0.108pt}}
\multiput(673.00,311.17)(4.132,-3.000){2}{\rule{0.450pt}{0.400pt}}
\put(679,307.17){\rule{1.300pt}{0.400pt}}
\multiput(679.00,308.17)(3.302,-2.000){2}{\rule{0.650pt}{0.400pt}}
\put(685,305.17){\rule{1.500pt}{0.400pt}}
\multiput(685.00,306.17)(3.887,-2.000){2}{\rule{0.750pt}{0.400pt}}
\multiput(692.00,303.95)(1.132,-0.447){3}{\rule{0.900pt}{0.108pt}}
\multiput(692.00,304.17)(4.132,-3.000){2}{\rule{0.450pt}{0.400pt}}
\put(698,300.17){\rule{1.300pt}{0.400pt}}
\multiput(698.00,301.17)(3.302,-2.000){2}{\rule{0.650pt}{0.400pt}}
\put(704,298.17){\rule{1.500pt}{0.400pt}}
\multiput(704.00,299.17)(3.887,-2.000){2}{\rule{0.750pt}{0.400pt}}
\put(711,296.17){\rule{1.300pt}{0.400pt}}
\multiput(711.00,297.17)(3.302,-2.000){2}{\rule{0.650pt}{0.400pt}}
\put(717,294.17){\rule{1.300pt}{0.400pt}}
\multiput(717.00,295.17)(3.302,-2.000){2}{\rule{0.650pt}{0.400pt}}
\multiput(723.00,292.95)(1.355,-0.447){3}{\rule{1.033pt}{0.108pt}}
\multiput(723.00,293.17)(4.855,-3.000){2}{\rule{0.517pt}{0.400pt}}
\put(730,289.17){\rule{1.300pt}{0.400pt}}
\multiput(730.00,290.17)(3.302,-2.000){2}{\rule{0.650pt}{0.400pt}}
\put(736,287.17){\rule{1.300pt}{0.400pt}}
\multiput(736.00,288.17)(3.302,-2.000){2}{\rule{0.650pt}{0.400pt}}
\put(742,285.17){\rule{1.300pt}{0.400pt}}
\multiput(742.00,286.17)(3.302,-2.000){2}{\rule{0.650pt}{0.400pt}}
\put(748,283.17){\rule{1.500pt}{0.400pt}}
\multiput(748.00,284.17)(3.887,-2.000){2}{\rule{0.750pt}{0.400pt}}
\put(755,281.17){\rule{1.300pt}{0.400pt}}
\multiput(755.00,282.17)(3.302,-2.000){2}{\rule{0.650pt}{0.400pt}}
\put(761,279.17){\rule{1.300pt}{0.400pt}}
\multiput(761.00,280.17)(3.302,-2.000){2}{\rule{0.650pt}{0.400pt}}
\put(767,277.17){\rule{1.500pt}{0.400pt}}
\multiput(767.00,278.17)(3.887,-2.000){2}{\rule{0.750pt}{0.400pt}}
\put(774,275.17){\rule{1.300pt}{0.400pt}}
\multiput(774.00,276.17)(3.302,-2.000){2}{\rule{0.650pt}{0.400pt}}
\put(780,273.17){\rule{1.300pt}{0.400pt}}
\multiput(780.00,274.17)(3.302,-2.000){2}{\rule{0.650pt}{0.400pt}}
\put(786,271.17){\rule{1.500pt}{0.400pt}}
\multiput(786.00,272.17)(3.887,-2.000){2}{\rule{0.750pt}{0.400pt}}
\put(793,269.17){\rule{1.300pt}{0.400pt}}
\multiput(793.00,270.17)(3.302,-2.000){2}{\rule{0.650pt}{0.400pt}}
\put(799,267.17){\rule{1.300pt}{0.400pt}}
\multiput(799.00,268.17)(3.302,-2.000){2}{\rule{0.650pt}{0.400pt}}
\sbox{\plotpoint}{\rule[-0.500pt]{1.000pt}{1.000pt}}%
\put(181.00,626.00){\usebox{\plotpoint}}
\put(196.16,611.84){\usebox{\plotpoint}}
\put(211.70,598.11){\usebox{\plotpoint}}
\put(227.11,584.24){\usebox{\plotpoint}}
\put(242.98,570.85){\usebox{\plotpoint}}
\put(259.31,558.07){\usebox{\plotpoint}}
\put(275.63,545.27){\usebox{\plotpoint}}
\put(291.89,532.41){\usebox{\plotpoint}}
\put(308.51,520.00){\usebox{\plotpoint}}
\put(325.62,508.26){\usebox{\plotpoint}}
\put(342.68,496.55){\usebox{\plotpoint}}
\put(359.74,484.84){\usebox{\plotpoint}}
\put(377.06,473.40){\usebox{\plotpoint}}
\put(394.57,462.28){\usebox{\plotpoint}}
\put(412.14,451.24){\usebox{\plotpoint}}
\put(430.07,440.95){\usebox{\plotpoint}}
\put(448.20,430.90){\usebox{\plotpoint}}
\put(466.40,421.07){\usebox{\plotpoint}}
\put(484.41,410.80){\usebox{\plotpoint}}
\put(503.15,401.90){\usebox{\plotpoint}}
\put(521.28,391.88){\usebox{\plotpoint}}
\put(540.03,382.99){\usebox{\plotpoint}}
\put(558.78,374.11){\usebox{\plotpoint}}
\put(577.53,365.23){\usebox{\plotpoint}}
\put(596.59,357.21){\usebox{\plotpoint}}
\put(615.64,349.18){\usebox{\plotpoint}}
\put(634.69,341.15){\usebox{\plotpoint}}
\put(653.74,333.13){\usebox{\plotpoint}}
\put(673.15,325.95){\usebox{\plotpoint}}
\put(692.54,318.73){\usebox{\plotpoint}}
\put(711.99,311.67){\usebox{\plotpoint}}
\put(731.41,304.53){\usebox{\plotpoint}}
\put(750.77,297.21){\usebox{\plotpoint}}
\put(770.57,290.98){\usebox{\plotpoint}}
\put(790.36,284.75){\usebox{\plotpoint}}
\put(805,280){\usebox{\plotpoint}}
\put(181.00,571.00){\usebox{\plotpoint}}
\put(196.82,557.65){\usebox{\plotpoint}}
\put(213.15,544.87){\usebox{\plotpoint}}
\put(229.10,531.58){\usebox{\plotpoint}}
\put(245.44,518.80){\usebox{\plotpoint}}
\put(262.19,506.68){\usebox{\plotpoint}}
\put(279.23,494.85){\usebox{\plotpoint}}
\put(296.09,482.81){\usebox{\plotpoint}}
\put(313.07,470.96){\usebox{\plotpoint}}
\put(330.63,459.92){\usebox{\plotpoint}}
\put(348.19,448.87){\usebox{\plotpoint}}
\put(366.12,438.59){\usebox{\plotpoint}}
\put(384.05,428.30){\usebox{\plotpoint}}
\put(401.99,418.01){\usebox{\plotpoint}}
\put(419.92,407.62){\usebox{\plotpoint}}
\put(438.07,397.62){\usebox{\plotpoint}}
\put(456.76,388.62){\usebox{\plotpoint}}
\put(475.06,378.97){\usebox{\plotpoint}}
\put(493.81,370.09){\usebox{\plotpoint}}
\put(512.78,361.74){\usebox{\plotpoint}}
\put(531.66,353.17){\usebox{\plotpoint}}
\put(550.62,344.79){\usebox{\plotpoint}}
\put(569.72,336.76){\usebox{\plotpoint}}
\put(588.90,329.05){\usebox{\plotpoint}}
\put(608.17,321.52){\usebox{\plotpoint}}
\put(627.35,313.71){\usebox{\plotpoint}}
\put(646.70,306.37){\usebox{\plotpoint}}
\put(666.04,298.99){\usebox{\plotpoint}}
\put(685.80,292.66){\usebox{\plotpoint}}
\put(705.31,285.63){\usebox{\plotpoint}}
\put(725.11,279.40){\usebox{\plotpoint}}
\put(744.86,273.05){\usebox{\plotpoint}}
\put(764.64,266.79){\usebox{\plotpoint}}
\put(784.43,260.52){\usebox{\plotpoint}}
\put(804.42,255.10){\usebox{\plotpoint}}
\put(805,255){\usebox{\plotpoint}}
\end{picture}
\caption{The pionium lifetime, in units of $10^{-15}$ s, as a function of 
the combination $(a_0^0-a_0^2)$ of the $S$-wave scattering lengths. The band 
delineated by the dotted lines takes account of the uncertainties, coming from 
theoretical evaluations, low energy constants and $a_1^1$. (Thanks to 
H.Sazdjian hep-ph/9911520).}
\label{fig:0}
\end{center}
\end{figure}
\vspace{-5mm}

The general expression for the pionium decay width
($\Gamma$), according to ChPT\cite{Gasser99} at leading and next-to-leading
order in isospin breaking, is
\begin{equation}\label{lifetime}
\frac{1}{\tau} = \Gamma = \frac{2}{9}\cdot \alpha^3 \cdot p^* \cdot
\left( a_0^0 - a_0^2 + \epsilon \right) ^2 \left( 1 + K \right) ,
\end{equation}
where~~~ $\alpha$ is the fine structure constant~;~\\
\indent \qquad $p^* = \left( M_{\pi^+}^2 - M_{\pi^0}^2 -
\alpha^2 M_{\pi^+}^2 /4\right)^{1/2}$ is the CM momentum of $\pi^0$~;~ \\
\indent \qquad $\epsilon = (0.58 \pm 0.16)\cdot 10^{-2}$~;
~$K=1.07\cdot 10^{-2}$.

Using Eq.(\ref{lifetime}), $a_0^0=0.206$ and $a_0^2=-0.0443$, Gasser et 
al.\cite{Gasser99} have evaluated the pionium lifetime in the ground state
$\tau = 3.25 \cdot 10^{-15} s$.
In the isospin symmetry limit $\epsilon = 0$;~$K = 0$, Eq.(\ref{lifetime})
becomes the Deser type formula\cite{Deser54}.

Eq.(\ref{lifetime}) allows to get a relative error of the scattering lengths 
difference
$\frac{\Delta \left( a_0^0 - a_0^2 \right)}{\left( a_0^0 - a_0^2 \right)}
\approx 5 \%$, if DIRAC measures the lifetime with a
$\frac{\Delta \tau}{\tau} \approx 10 \%$ error.

According to GChPT\cite{Sazdjian99} the $\tau$ vs. ($a_0^0 - a_0^2$) dependence 
is presented in Fig.\ref{fig:0}. Hence the experimental determination of the 
pionium lifetime could be interpreted in quark condensation terms for three 
different cases: First, if the central value of $\tau$ is close to 
$3\times 10^{-15}~s$, lying above 
$2.9 \times 10^{-15}~s$, then the strong condensate assumption of ChPT is 
firmly confirmed, since its predictions of ($a_0^0 - a_0^2$) lie between 
0.250 and 0.258. Second, if the central value of $\tau$ lies below 
$2.4\times 10^{-15}~s$, it is the scheme of GChPT, which is confirmed, since 
the corresponding central value of $\alpha$ would lie above 2. The third 
possibility is the most difficult to interpret. If the central value of $\tau$
lies in the interval $2.4 \div 2.9 \times 10^{-15}~s$, then, because of the
uncertainties, ambiguities in the interpretation may arise.

\section{Experimental method}
In an abundent production of oppositely charged pions the Coulomb interaction
can form atomic $\pi^+ \pi^-$ bound states. If the pions have a small relative 
momentum in their CM system ($Q \sim 1~MeV/c$) and are much closer than the 
Bohr radius ($387~fm$), then pionium atoms are produced with a high production 
probability due to the large wave function overlap. Such pions originate from 
short-lived sources ($\rho, \omega, \Delta$), and not from long-lived ones 
($\eta, K_s^0$), because in the latter case the separation of the two pions is 
in most cases larger than the Bohr radius.

The pionium production cross section is proportional to the double inclusive
cross section $d\sigma_s^0 / (d \vec p_1 d \vec p_2)$ for $\pi^+ \pi^-$ pairs
from short-lived sources, without Coulomb interaction in the final 
state\cite{Nemenov85}, and to the squared atomic wave function of $nS$-states
at the origin $\left| \psi_n(0)\right|^2$
\begin{equation}\label{cross}
\frac{d \sigma_n^A}{d \vec p_A} = (2 \pi)^3 \frac{E_A}{M_A} 
\left|\psi_n(0)\right|^2 \frac{d\sigma_s^0}{d\vec p_1 
d\vec p_2}_{\left|\vec p_1=\vec p_2=\frac{\vec p_A}{2} \right.}
\end{equation}
where $\vec p_A, E_A$ and $M_A$ are momentum, energy and mass of the pionium 
atom in the Lab system respectively; $\vec p_1$ and $\vec p_2$ are the $\pi^+$ 
and $\pi^-$ momenta in the Lab system, and they must satisfy the relation 
$\vec p_1 = \vec p_2 = \vec p_A/2$ to form the atomic bound state.
Atoms formed in this way are in a $S$-state.

\begin{figure}[h]
\begin{center}
\epsfxsize=75mm 
\epsfbox{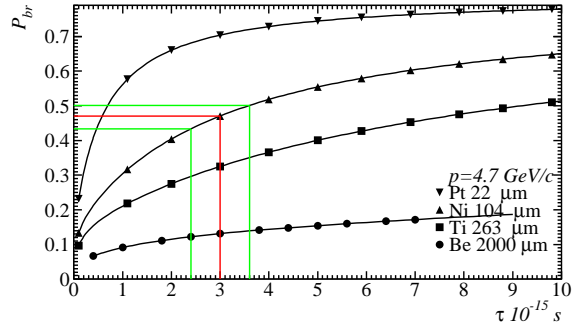} 
\caption{Probability of pionium break-up in the target.}
\label{fig:1}
\end{center}
\vspace{-3mm}
\end{figure}

After production in hadron-nucleus interaction relativistic pionium atoms
($2~GeV/c < p_A < 6~GeV/c$) are moving in the target. They can decay or, due to
electromagnetic interaction with the target material, get excited or
broken-up (ionized). Using atomic interaction cross sections, for a given target
material and thickness, one can calculate the break-up probability for
arbitrary values of pionium momentum and lifetime. In Fig.\ref{fig:1} there
are presented these dependencies for the pionium momentum $p=4.7~GeV/c$. 

Comparison of the measured break-up probability $P_{br} = n_A / N_A$ (ratio of 
broken-up - $n_A$ and produced - $N_A$ pionium atoms) with the calculated 
dependence of $P_{br}$ on $\tau$ (see Fig.\ref{fig:1}) gives a value of the 
lifetime.

The break-up process gives characteristic $\pi^+ \pi^-$ pairs, called 
{\sl atomic pairs}. They have a small relative momentum in their CM system 
($Q < 3~MeV/c$), a small opening angle 
($\theta_\pm \approx 6/\gamma \approx 0.35~mrad$ for $p_A = 4.7~GeV/c$) and 
nearly identical energies in the Lab system ($E_+ = E_-$ at 0.3 \% level).

\subsection{Determination of broken-up pionium atoms ($n_A$)}
The measurement of broken-up $n_A$ pionium atoms is realized through the 
analysis of the experimental distribution in $Q$ of $\pi^+ \pi^-$ pairs.

The free pion pair distribution can be written as the sum of the non-Coulomb 
(nC) (no final state interaction) and the Coulomb (C) pair distribution 
(Fig.\ref{fig:2}):
\begin{figure}[h]
\begin{center}
\begin{tabular}{cc}
\epsfxsize=30mm
\epsfbox{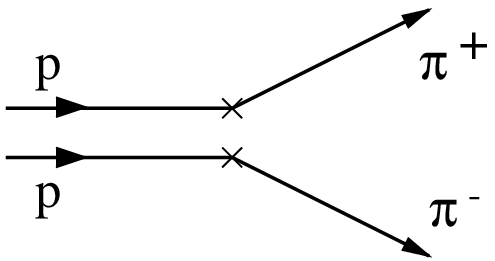} 
&
\epsfxsize=30mm
\epsfbox{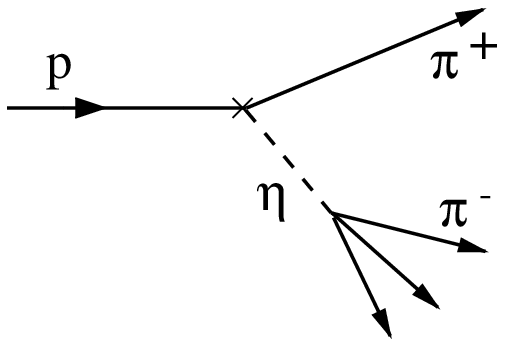} 
\end{tabular}
\begin{tabular}{cc}
\epsfxsize=30mm
\epsfbox{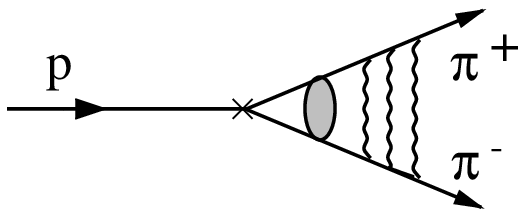} 
&
\epsfxsize=30mm
\epsfbox{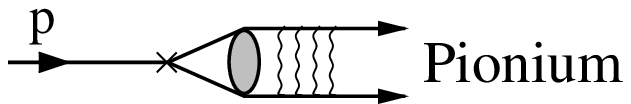} 
\end{tabular}
\caption{Accidental, non-Coulomb (long-lived sources) and Coulomb (short-lived 
sources) pion pair production and pionium production}
\label{fig:2}
\end{center}
\vspace{-5mm}
\end{figure}
$$
\frac{d N^{free}}{dQ} = \frac{d N^{nC}}{dQ} + \frac{d N^{C}}{dQ}
$$
$$
\frac{d N^{nC}}{dQ} \sim \frac{d N_{acc}^{exp}}{d Q} \equiv \Phi (Q) 
\qquad ; \qquad
\frac{dN^{C}}{dQ} \sim \Phi(Q) A_c(Q) (1 + a Q)
$$
where $A_c(Q)$ is the Coulomb and $(1+a Q)$ is the strong correlation factors.

Here we assumed that the non-Coulomb distribution of $\pi^+ \pi^-$ 
pairs (without FSI) can be extracted from the experimental distribution of 
accidental pairs $\Phi(Q)$. The free pion pair distribution is then given by
\begin{equation}\label{free}
\frac{d N^{free}}{dQ} = \frac{d N^{nC}}{dQ} + \frac{d N^{C}}{dQ} =
N_0 \Phi(Q) \left[ f + A_c(Q) (1 + a Q) \right],
\end{equation}
$N_0, f, a$ - free parameters. In the region $Q>3~MeV/c$, there are mainly free 
pairs. This part of the distribution is fitted with the function (\ref{free})
containing the experimental distribution $\Phi(Q)$ of $\pi^+ \pi^-$ accidental
pairs. The extrapolation of the approximation function to the region $Q<2~MeV/c$
yields the number of free pairs in this region. Hence the value $n_A$ of atomic 
pairs is 
\begin{equation}\label{nA}
n_A = \int \limits_{Q<2} \left( \frac{dN^{exp}}{dQ} - \frac{d N^{free}}{dQ}
\right) dQ . 
\end{equation}
From the measured break-up probability $P_{br} = n_A/N_A$, where $N_A$ is the
calculated total number (according to (\ref{cross})) of produced pionium
atoms, and from the dependence of $P_{br}$ on the lifetime $\tau$, one 
can derive a pionium ground state lifetime and hence a value for 
$\Delta = \left| a_0^0 - a_0^2 \right|$.

\section{Experimental setup}
\begin{figure}[h]
\vspace{-18mm}
\begin{center}
\epsfxsize=9cm
\epsfbox{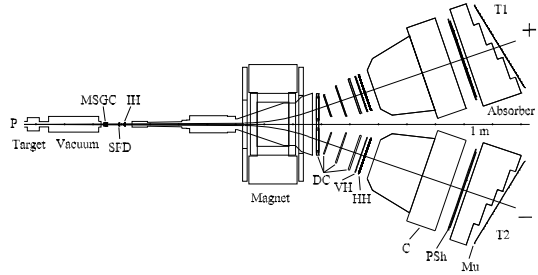}
\caption{Experimental setup}
\label{fig:3}
\end{center}
\vspace{-8mm}
\end{figure}

The experimental setup\cite{Lanaro99} (Fig.\ref{fig:3}) has been designed to 
detect pion pairs and to select atomic pairs at low relative momentum 
with a resolution better than 1~MeV/c. It was installed and commissioned in 
1998 at the ZT8 beam area of the PS East Hall at CERN. After a calibration run 
in 1998, DIRAC has been collecting data since summer 1999.

The 24~GeV/c proton beam extracted from PS is focused on the target. The
secondary particle channel, with an aperture of 1.2 msr, has the reaction
plane tilted upwards at $5.7^\circ$ relative to the horizontal plane. It 
consists of the following components: 
4 planes of Micro Strip Gas Chambers (MSGC) with $4\times 512$ channels; 
2 planes Scintillation Fiber Detector (SciFi) with $2\times 240$ channels; 
2 planes Ionization Hodoscope (IH) with $2\times 16$ channels; 
1 Spectrometer Magnet of $2.3~Tm$ bending power. 
Downstream to the magnet the setup splits into two arms placed at 
$\pm 19^\circ$, relative to the central axis. Each arm is equipped with 
a set of identical detectors: 
4 Drift Chambers (DC), the first one common to both arms and with 6 planes and
800 channels, the other DC's have altogether per arm 8 planes and 608
channels; 
1 Vertical scintillation Hodoscope (VH) plane with 18 channels; 
1 Horizontal scintillation Hodoscope (HH) plane with 16 channels; 
1 Cherenkov detectors (Ch) with 10 channels; 
1 Preshower scintillation detector (PSh) plane with 8 channels; 
1 Muon counter (Mu) plane with (28+8) channels. 

For suppressing the large background rate a multilevel trigger was designed to
select atomic pion pairs. The trigger levels are defined as follows: 
$T_0 = (VH \cdot PSh)_1 \cdot (VH \cdot PSh)_2 \cdot IH$, fast zero level 
trigger; 
$T_1 = (VH \cdot HH \cdot \overline{Ch}\cdot PSh)_1 \cdot 
(VH \cdot HH \cdot \overline{Ch}\cdot PSh)_2$, first level trigger from the 
downstream detectors; 
$T_2 = T_0 \cdot (IH \cdot SciFi)$, second level trigger from the upstream 
detectors, which selects particle pairs with small relative distance; 
$T_3$ is a logical trigger which applies a cut to the relative momentum of 
particle pairs. It handles the patterns of VH and IH detectors. $T_3$ did not
so far trigger the DAQ system, but its decisions were recorded.

An incoming flux of $\sim 10^{11}$ protons/s would produce a rate of 
secondaries of about $3\times 10^6$/s in the upstream detectors and 
$1.5 \times 10^6$/s in the downstream detectors. At the trigger level this rate 
is reduced to about $2\times 10^3$/s, with an average event size of about 0.75
Kbytes.

With the $95~\mu m$ thin $Ni$ target, the expected average pionium yield within 
the setup acceptance is $\sim 0.7 \times 10^{-3}$/s, equivalent to a total 
number of $\sim 10^{13}$ protons on target to produce one pionium atom.

\section{First experimental results}

The data taking has been done mainly with $\pi^+ \pi^-$ and $p \pi^-$ pairs and 
also $e^+ e^-$ pairs for detector calibration. For the first data analysis only 
the most simple events were selected and processed, those with a single track 
in each arm, with signals in DC, VH and HH. The tracks in the DC's were 
extrapolated to the target plane crossing point of the proton beam. A cut was 
applied along $X$ and $Y$ distances between the extrapolated track and the hit 
fiber of the SciFi planes ($18 mm$ divided by the particle momentum in GeV/c, 
to take into account the multiple scattering effect). Finally, these events 
were interpreted as $\pi^+ \pi^-$ or $p \pi^-$ pairs produced in the target.

\begin{figure}[h]
\vspace{-15mm}
\begin{center}
\begin{minipage}[t]{65mm}
\mbox{\epsfig{figure=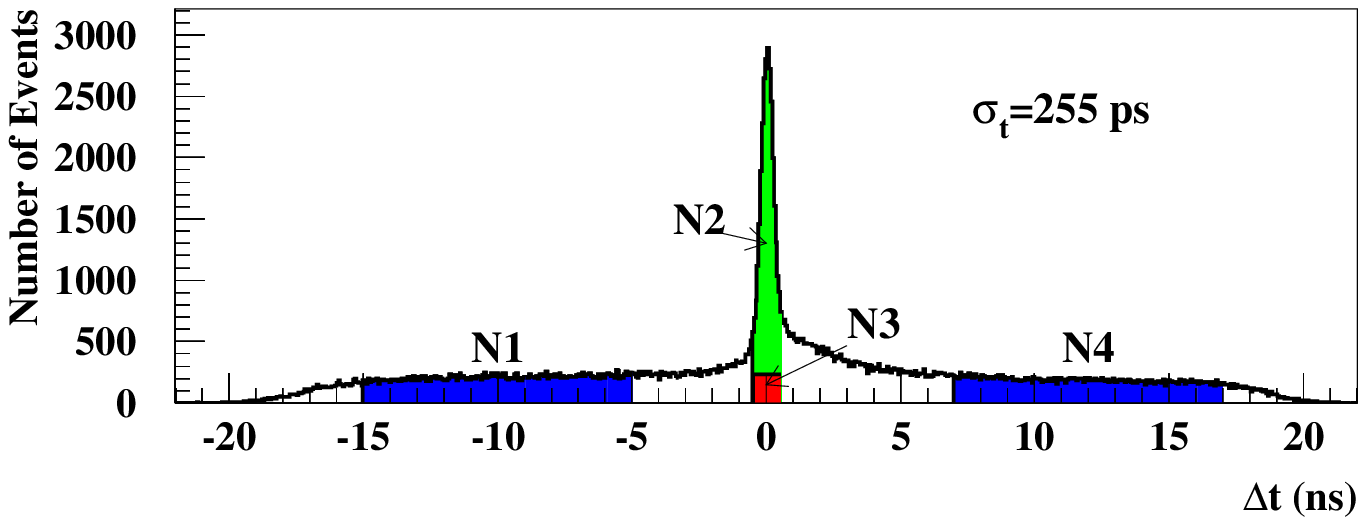,width=65mm}}
\caption{VH time-of-flight difference distribution for pair events}
\label{fig:4}
\end{minipage}
~
\begin{minipage}[t]{45mm}
\mbox{\epsfig{figure=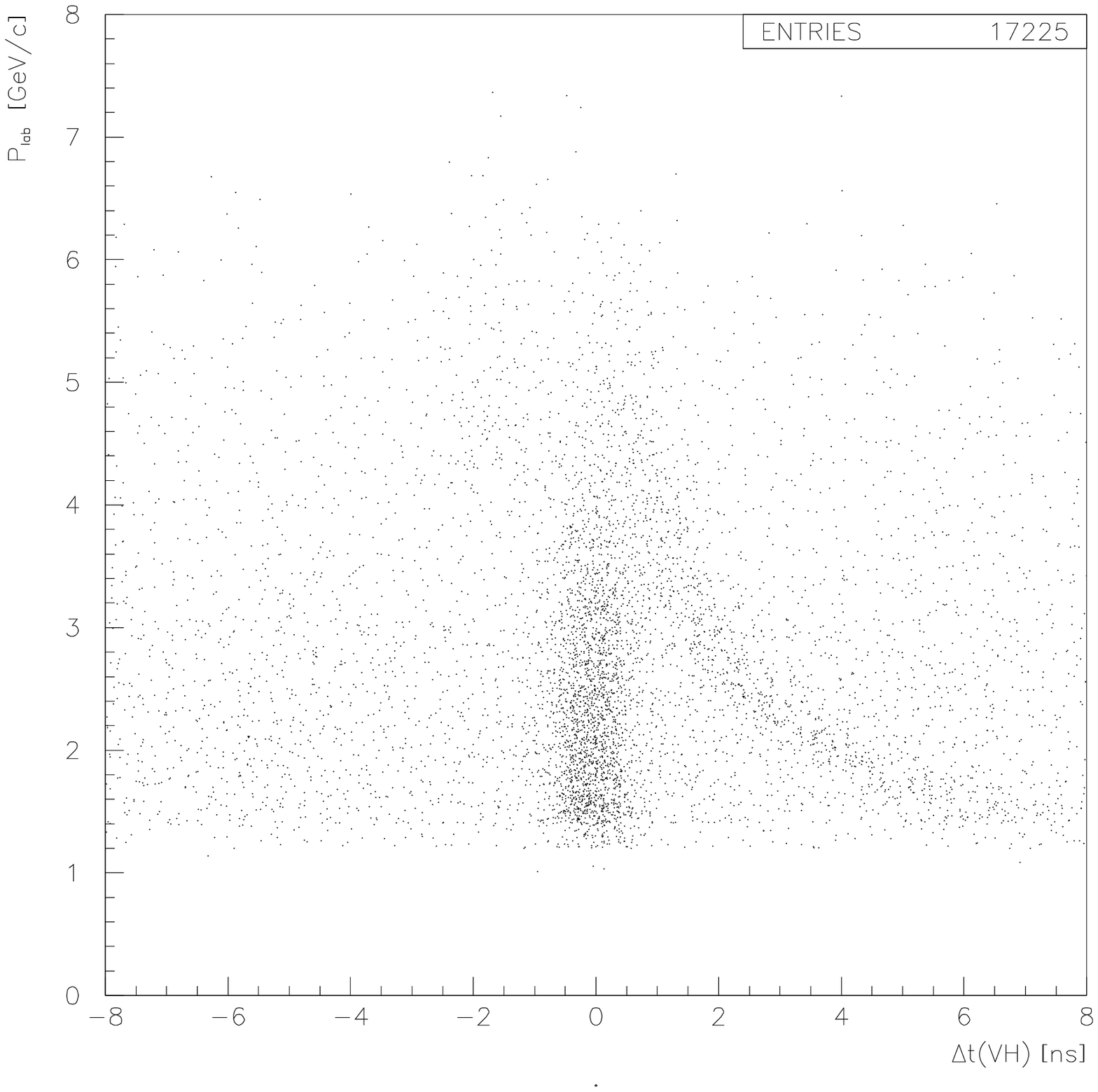,width=45mm}}
\caption{Positive particle momentum versus VH time-of-flight difference for 
particle pairs}
\label{fig:5}
\end{minipage}
\end{center}
\vspace{-5mm}
\end{figure}

The difference in the time-of-flight $\Delta t$ between the positive 
particle (left arm) and negative particle (right arm) of the pair at the level 
of VH is presented in Fig.\ref{fig:4}. 

The first interval $-20 <\Delta t < -0.5~ns$ corresponds to accidental hadron 
pairs (mainly $\pi ^+ \pi ^-$). In the second interval $-0.5<\Delta t< 0.5~ns$ 
one observes the peak of coincidence hits associated to correlated hadron pairs 
over the background of accidental pairs. The width of the correlated pair peak 
yields the time resolution of the VH ($\sigma_t \approx 250~ps$). The asymmetry
on the right side of the peak is due to admixture of protons in the $\pi^+$ 
sample, that are $p \pi^-$ events. Hence the third interval 
$0.5 < \Delta t < 20~ns$ contains both accidental pairs and $p \pi^-$ events.

This time-of-flight discrimination between $\pi^+ \pi^-$ and $p \pi^-$ events
is effective for momenta of positive particles below 4.5~GeV/c. This is
demonstrated 
in Fig.\ref{fig:5}, where the scatter plot of positive particle momentum versus
difference in time-of-flight $\Delta t$ in VH is shown. The single particle 
momentum interval accepted by spectrometer is $1.3 \div 7.0$~GeV/c.

For correlated $\pi^+ \pi^-$ pairs Coulomb interaction in the final state has 
to be considerated, because it increases noticeably the yield of $\pi^+ \pi^-$ 
pairs with low relative momentum in CM ($Q < 5~MeV/c$). For accidental pairs 
this enhancement is absent. 
\begin{figure}[h]
\vspace{-2mm}
\begin{center}
\begin{minipage}[t]{57mm}
\mbox{\epsfig{figure=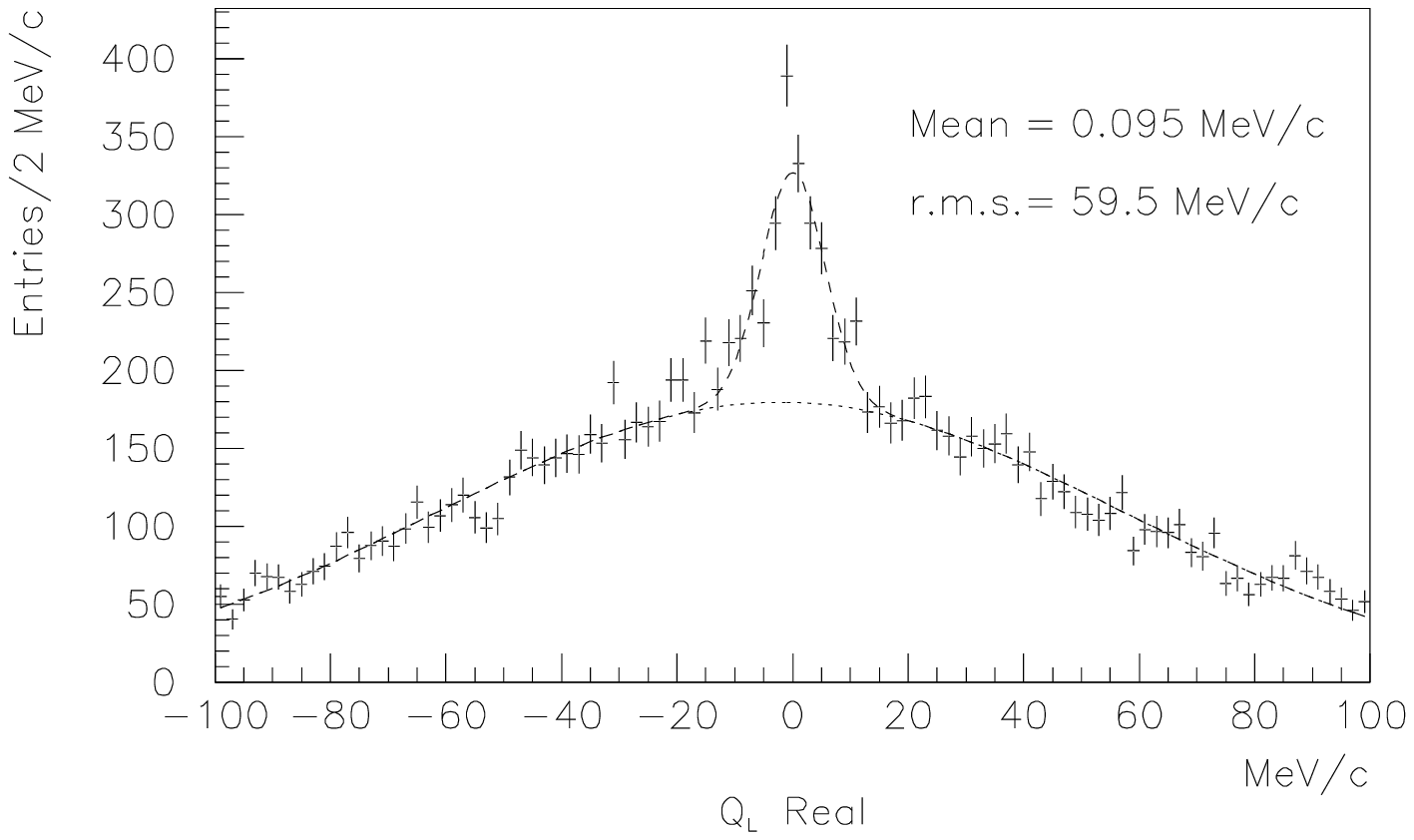,width=57mm}}
\caption{Correlated $\pi^+ \pi^-$ pairs with po\-sitive particle momenta 
$p_{lab} <~ 4.5$~GeV/c and $Q_{T} < ~4$~MeV/c.}
\label{fig:6}
\end{minipage}
~
\begin{minipage}[t]{57mm}
\mbox{\epsfig{figure=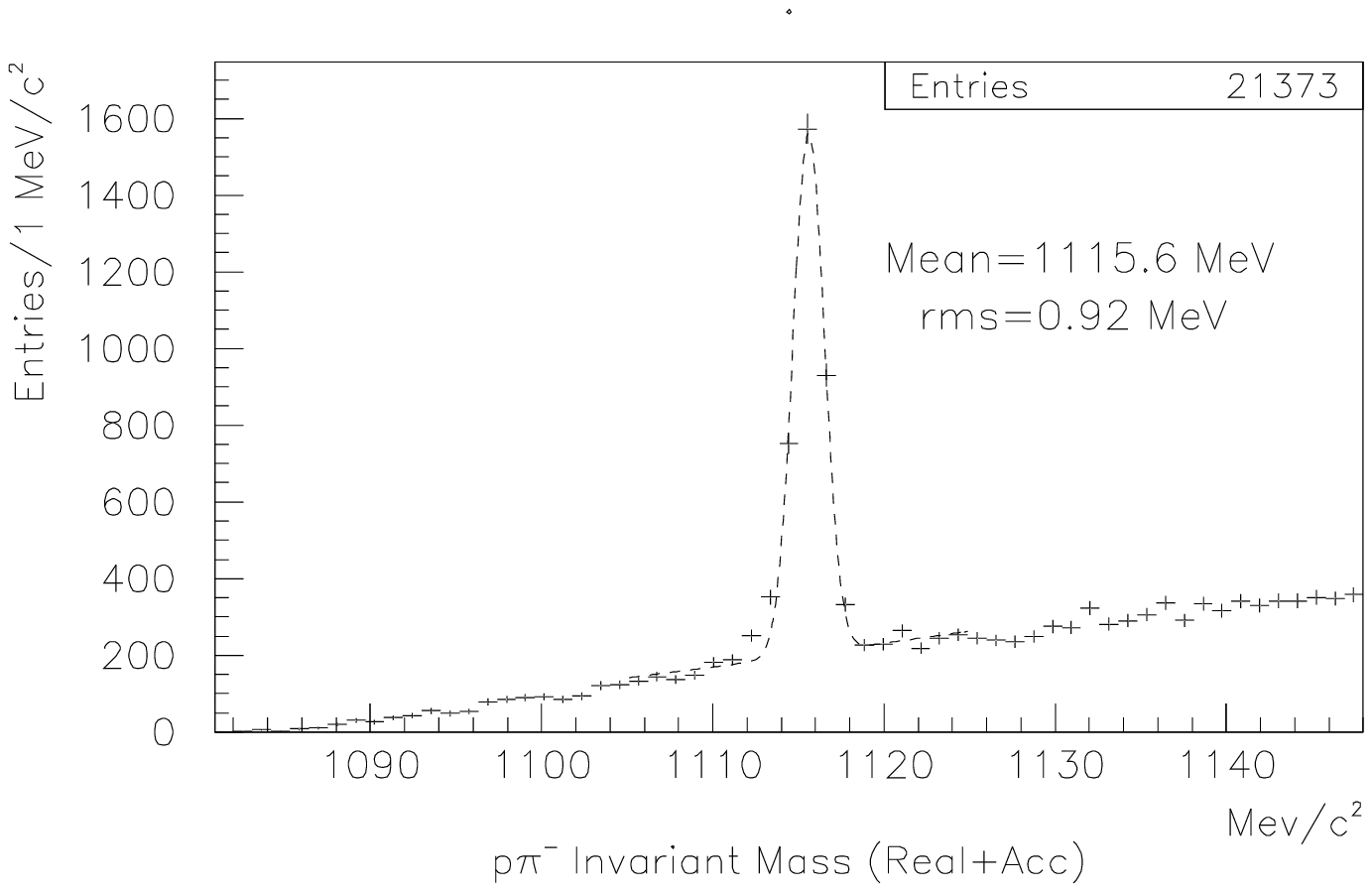,width=57mm}}
\caption{$p \pi^-$ invariant mass for proton momenta $p_{lab} >~ 3$~GeV/c.}
\label{fig:7}
\end{minipage}
\end{center}
\vspace{-5mm}
\end{figure}

Fig.\ref{fig:6} shows the distribution of the longitudinal component $Q_L$ (the 
projection of $Q$ along the total momentum of the pair) for correlated pairs. 
There are plotted pair events with positive particle momentum 
$p_{lab} < 4.5~GeV/c$, occuring within the "correlated" $\Delta t$ peak and 
with transversal component $Q_T < 4~MeV/c$ to increase the fraction of low 
relative momentum pairs.

In the region $|Q_L| \leq 10~MeV/c$ there is a noticeable enhancement of
correlated $\pi^+ \pi^-$ pairs due to Coulomb attraction in the final state.

The most important parameter for data analysis is the resolution in $Q_L$ and 
$Q_T$. This has been measured by the reconstruction of the invariant mass of 
$p \pi^-$ pairs. The distribution of $p \pi^-$ invariant mass is presented in 
Fig.\ref{fig:7}. Positive particles are restricted to momenta larger than 3~ 
GeV/c, and the time-of-flight must lie in $0.5 < \Delta t < 18~ns$. A clear 
peak at the $\Lambda$ mass $m_\Lambda = 1115.6~MeV/c^2$ with a standard 
deviation $\sigma = 0.92~MeV/c^2$ can be seen. These mass parameter values 
show a good detector calibration and coordinate detector alignment, with 
an accuracy in momentum reconstruction better than 0.5 \% in the kinematic 
range of $\Lambda$ decay products. This gives for the relative momentum 
resolution $\sigma_Q \sim 2.7~MeV/c$. For $\pi^+ \pi^-$ pairs a better 
resolution can be obtained, due to the different kinematics.
\vspace{-3mm}
\section{Conclusion}
The DIRAC setup test and calibration have been done, and the DIRAC expe\-riment
began data taking. To achieve the goal - measurement of the pionium lifetime
with 10\% precision - we have to consider a number of at least 20000 recorded
"atomic pairs". Improvements in hardware and software will continue this year.
These will result in a better data quality.

\end{document}